\documentclass[10pt, final, journal, letterpaper, twocolumn]{IEEEtran}
\ifCLASSINFOpdf
\else
\fi
\makeatletter
\def\ps@headings{%
\def\@oddhead{\mbox{}\scriptsize\rightmark \hfil \thepage}%
\def\@evenhead{\scriptsize\thepage \hfil \leftmark\mbox{}}%
\def\@oddfoot{}%
\def\@evenfoot{}}
\makeatother 

\IEEEoverridecommandlockouts
\usepackage{amsmath}  
\usepackage{color}
\usepackage{amsfonts}
\usepackage[dvips]{graphicx}
\usepackage{caption}
\usepackage{subfigure}
\usepackage{times}
\usepackage{cite}
\usepackage{lettrine}
\usepackage{amsmath}
\allowdisplaybreaks[4]
\usepackage{array}
\usepackage{amssymb}

\usepackage{stfloats}
\usepackage{graphicx}
\usepackage{footnote}
\usepackage{booktabs}
\usepackage{array}
\usepackage{algorithmic}
\usepackage{algorithm}
\usepackage{subeqnarray}
\usepackage{cases}
\usepackage{threeparttable}
\usepackage{color}
\usepackage{url}
\usepackage{bm}

\newtheorem{theorem}{\underline{Theorem}}[section]
\newtheorem{lemma}{\underline{Lemma}}[section]

\begin{document}
\bibliographystyle{IEEEtran}

\title{Energy Efficiency of Distributed Antenna Systems with Wireless Power Transfer}
\IEEEoverridecommandlockouts

\author{\IEEEauthorblockN{Yuwen Huang, Yuan Liu,~\IEEEmembership{Senior Member,~IEEE}, and Geoffrey Ye Li,~\IEEEmembership{Fellow,~IEEE}}

%}

%\author{\IEEEauthorblockN{Yuwen Huang and Yuan Liu}
%\vspace*{0.5em}
%\IEEEauthorblockA{School of Electronic and Information Engineering,
%South China University of Technology, Guangzhou 510641, China}
%
%\IEEEauthorblockA{
%Emails: eehyw@mail.scut.edu.cn, eeyliu@scut.edu.cn}
%
%}
%
\thanks{
Manuscript received March 18, 2018; revised July 2, 2018; accepted September 6, 2018. This work was supported in part by the Natural Science Foundation of China under Grant 61771203 and Grant U1701265, and in part by the Pearl River Science and Technology Nova Program of Guangzhou under Grant 201710010111. \emph{(Corresponding author: Yuan Liu.)}

Y. Huang and Y. Liu are with School of Electronic and Information Engineering, South China University of Technology, Guangzhou 510641, China (email: eehyw@mail.scut.edu.cn, eeyliu@scut.edu.cn). 

G. Y. Li is with the School of Electrical and Computer Engineering, Georgia Institute of Technology, Atlanta, GA 30332-0250 USA (e-mail: liye@ece.gatech.edu).}
}
\maketitle

%\IEEEauthorblockA{\IEEEauthorrefmark{2} National Mobile Communications Research Laboratory, Southeast University, Nanjing 210096, China
%}

%\author{Yuwen Huang,~\IEEEmembership{Student}, and Yuan Liu,~\IEEEmembership{Member,~IEEE}
%\thanks{Copyright (c) 2013 IEEE. Personal use of this material is permitted. However, permission to use this material for any other purposes must be obtained from the IEEE by sending a request to pubs-permissions@ieee.org.}
%\thanks{Manuscript received November 20, 2014; revised January 31, 2015, April 22, 2015, June 30, 2015, and August 10, 2015; accepted September 14, 2015. This work is supported in part by the Natural Science Foundation of China under Grant 61401159, the open research fund of National Mobile Communications Research Laboratory, Southeast University, under Grant 2016D06, and the Fundamental Research Funds for the Central Universities.
%%
%The associate editor coordinating the review of this manuscript and approving it for publication was Prof. K. Ren.}

%\thanks{Y. Huang is with School of Electronic and Information Engineering,
%South China University of Technology, Guangzhou, 510641, P. R. China. Email: eehyw@mail.scut.edu.cn. }
%
%\thanks{Y. Liu \emph{(Corresponding Author)} is with School of Electronic and Information Engineering,
%South China University of Technology, Guangzhou, 510641, P. R. China, and also with National Mobile Communications Research Laboratory, Southeast University, P. R. China. Email: eeyliu@scut.edu.cn.}
%}
\vspace{-1.5cm}
\maketitle
\begin{abstract}
In this paper, we study energy-efficient resource allocation in distributed antenna system (DAS) with wireless power transfer, where time-division multiple access (TDMA) is adopted for  downlink multiuser information transmission. In particular, when a user is scheduled to receive information, other users harvest energy at the same time using the same radio-frequency (RF) signal. We consider two types of energy efficiency (EE) metrics: user-centric EE (UC-EE) and network-centric EE (NC-EE). Our goal is to maximize the UC-EE and NC-EE, respectively, by optimizing the transmission time and power subject to the energy harvesting requirements of the users. For both UC-EE and NC-EE maximization problems, we transform the nonconvex problems into equivalently tractable problems by using suitable mathematical tools and then develop iterative algorithms to find the globally optimal solutions. Simulation results demonstrate the superiority of the proposed methods compared with the benchmark schemes.  
\end{abstract}


 \section{Introduction}
The next generation wireless communication systems are expected to provide $1000\times$ increase in data traffic and support billions of internet-of-things (IoT) devices. However, the limitation of battery capacity will be a bottleneck and it is of vital importance to prolong the lifetime of energy-constrained wireless devices. To this end, wireless information and power transfer (WIPT) has been regarded as a promising solution to achieve two-way communications and at the same time provide cost-effective energy supplies for low-power IoT devices. Rather than relying solely on the batteries, IoT devices are also able to replenish energy by WIPT in a sustainable and controllable way \cite{Lu2015}. In general, there are two solutions to implement radio-frequency (RF) based WIPT in practice: wireless powered communication networks (WPCNs) and simultaneous wireless information and power transfer (SWIPT).  In WPCNs, wireless nodes are first powered by an energy transmitter and then use the harvested energy to transmit data \cite{Ju2014c,Clerckx2017,Ju2014a}, while SWIPT uses the same RF signal to convey energy and information simultaneously \cite{Liu2016,Zhang2013,Liu2016a,Liu2013,Zhang2016c,Liu2017a,Clerckx2018,Liu2017}.

However, WIPT suffers from the fast decay of wireless energy transmission over distances.  The traditional way to deal with this problem is energy beamforming using multi-antenna techniques, which is also not efficient due to the distance limitation of WIPT. Interestingly, this problem can be alleviated in distributed antenna system (DAS) \cite{Zhou2003,hu2007,He2013,Li2016}. Different from the conventional base stations with co-located antennas, the role of the base station in DAS is substituted by a central processor (CP) and a set of distributed antenna (DA) ports. Specifically, the CP is designed for the computational intensive baseband signal processing and the DA ports, geographically distributed throughout the area and connected to the CP via high capacity backhaul links, are used for all RF signal's operations. Thus DAS can substantially improve system's coverage and throughput. More importantly, as the access distances between the users and the DA ports are substantially reduced in DAS, WIPT is more flexible and efficient. As a result, many studies have been made to integrate WIPT into DAS.  For instance, the security of WIPT based DAS has been investigated in \cite{Ng2015}. WIPT in massive DAS has been considered in \cite{Yuan2015}, while SWIPT for multiple-input single-output (MISO) DAS has been investigated in \cite{Yuan2017}.
%He2012,He2014,He2015,Kim2015,Ren2017

 On the other hand, due to the rapidly increasing energy cost in communication systems, energy efficiency (EE) has been considered as an important system performance metric \cite{Zhang2017,Miao2013}. EE optimization has been widely studied in WIPT \cite{Ng2013a,Xiong2014,Ng2013,Huang2018a} from a network-centric (NC) perspective, namely, NC-EE. As shown in \cite{Huang2018a}, the NC-EE of WIPT usually leads to unbalanced or unfair energy consumption, i.e., some users consume most of the network resources while others may be idle. Such a NC-EE optimization is optimal for overall system design. Besides that, improving the EEs of individual users is equally important for improving users' qualities of experience (QoE), because users have different battery capacities and heterogeneous QoE requirements. Therefore, some works \cite{Yu2015,Yu2015a,Wu2016b,Ding2018}  have considered weighted sum EEs of individual users as the performance metric, i.e., user-centric EE (UC-EE). For example, joint downlink and uplink resource allocation for UC-EE maximization has been studied in \cite{Yu2015}. UC-EE in multiple radio access technologies (RATs) heterogeneous network (HetNet) has been considered in \cite{Yu2015a}. Joint time allocation and power control has been studied in WPCNs for UC-EE maximization in \cite{Wu2016b}, where the users first harvest energy from a dedicated energy station and then transmit information to an access point using the harvested energy in time-division multiplexing access (TDMA) manner. UC-EE in WPCNs has also been investigated in \cite{Ding2018}, where users are allowed to transmit information simultaneously in the uplink channel. 

In this paper, we study both UC-EE and NC-EE in WIPT based DAS as shown in Fig. \ref{fig:system}, where TDMA is adopted for downlink multiuser information transmission. In the considered system, when a user is scheduled for receiving information, such as user 2 in the figure, the remaining users harvest energy from the same RF signal, such as user 1, at the same time. Different from \cite{Wu2016b,Ding2018}, in the considered system there is no need for extra dedicated energy signal to charge the users, and each user harvests energy when it is not scheduled for receiving information. Additionally, rather than being served by a single energy station and a single access point in \cite{Wu2016b,Ding2018}, in the considered system each user can receive information from a different group of DA ports due to the geographical distribution, so that the DA ports for information decoding and energy harvesting for each user may be different, making the system more flexible.   
%There are many reasons for considering UC-EE in WIPT based DAS. First, in WIPT based DAS, users are able to harvest energy and decode information independently. Normally they are serviced by a different set of DA ports. So this system can be viewed as a user-centric system in some way. In such a user-centric system, each user's experience, i.e., UC-EE ought to be highly regarded. Furthermore, rather than considering NC-EE, considering UC-EE in WIPT based DAS can present some interesting insights such as the performance trade-off between users. For example, if a user has strict minimum harvested energy constraint and a short period of time for energy harvesting (EH), it usually receives higher transmit power during the time for EH when other users perform information decoding (ID) alternatively, leading to the poor users' EEs performance among other users. As for the case that this user has a long period of time for EH, other users will have a high flexibility in allocating the transmit power from DA ports, which facilitates better users' EEs performance. However, the optimization of NC-EE can not reflect this phenomenon where we have to balance each user's EE and take their individual experience into consideration. This process is meaningful for the user-centric application. In addition, in UC-EE maximization problem we have the weight factor which reflects the priorities between users and gives us the flexibility to customize the experience of different users.

The main contributions of this paper are summarized as follows: 
\begin{itemize}
	\item We study energy-efficient resource allocation in WIPT based DAS for both UC-EE and NC-EE maximization. Each un-scheduled user is allowed to harvest energy from the information bearing signals conveyed for other users. We jointly optimize transmit time and power for TDMA-based multiuser transmission while satisfying the minimum harvested energy requirement for each user and the maximum transmit power budget for each DA port. 
	\item For the UC-EE maximization problem, the objective function has the structure of  the sum-of-ratios. Therefore,  we convert it into an equivalent subtractive form by introducing a set of auxiliary parameters, and then propose an iterative algorithm to solve the equivalent optimization problem in two layers. In the inner layer, the subtractive formed problem is optimally solved by Lagrangian duality method because of the concavity of the transformed problem. In the outer layer, we update the auxiliary parameters with the damped Newton method, which ensures global convergence to the optimal solution of the original problem. 	
	\item We also investigate the NC-EE maximization problem in the same system, which is a fractional programming problem. We also develop a two-layer iterative algorithm to find the optimal solution. In the inner layer, two block coordinate descent (BCD) optimization loops are proposed to find the optimal time and power allocation. In the outer layer, the Dinkelbach method is used for solving the fractional structure of the original problem.  
 \end{itemize}

The rest of this paper is organized as follows. Section \ref{se1} introduces the system model and formulates the UC-EE maximization problem and NC-EE maximization problem, respectively. Sections \ref{se2} and \ref{se4} solve the UC-EE and NC-EE maximization problems, respectively. Section \ref{se5} provides extensive simulation results and discussions. Finally, Section \ref{se6} concludes the paper.
%Section \ref{se3} details the efficient suboptimal algorithm for UC-EE maximization problem. %
 \section{System Model and Problem Formulation}\label{se1}
In this section, we first introduce the system model of the WIPT based DAS. Then we formulate the UC-EE and NC-EE maximization problems, respectively.
\subsection{System Model}
\begin{figure}[t]
\begin{centering}
\includegraphics[scale=0.6]{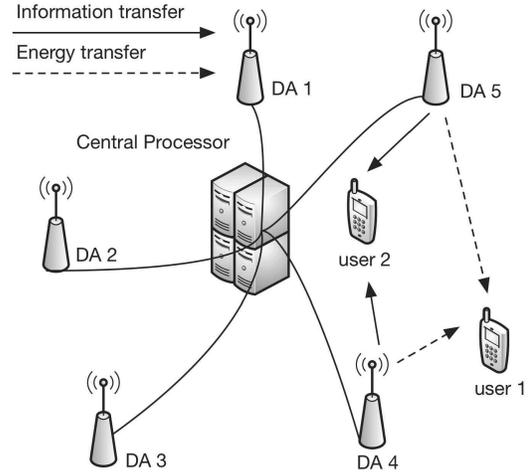}
\vspace{-0.1cm}
 \caption{ An example of system model of DAS. }\label{fig:system}
\end{centering}
\vspace{-0.1cm}
\end{figure}

%\begin{figure}
%\begin{minipage}[t]{0.5\linewidth}
%\centering
%\includegraphics[scale=0.4]{Single-user.eps}
%\caption{Structure of DAS with three distributed antenna ports (N=3).}
%\label{fig:system}
%\end{minipage}%
%\begin{minipage}[t]{0.5\linewidth}
%\centering
%\includegraphics[scale=0.5]{Receiver-structure.eps}
%\caption{ Receiver's structure.}
%\label{fig:receiver}
%\end{minipage}
%\end{figure}

%\begin{figure}[t]
%\begin{centering}
%\includegraphics[scale=0.6]{Single-user.eps}
%\vspace{-0.1cm}
% \caption{ An example of system model of DAS with five DA ports. }\label{fig:system}
%\end{centering}
%\vspace{-0.1cm}
%\end{figure}
%\begin{figure}[t]
%\begin{centering}
%\includegraphics[scale=0.6]{IoT_structure.eps}
%\vspace{-0.1cm}
% \caption{ The receiver circuit of the SWIPT based IoT devices. }\label{fig:receiver}
%\end{centering}
%\vspace{-0.1cm}
%\end{figure}

As shown in Fig. \ref{fig:system}, we consider a WIPT based downlink DAS consisting of a CP, $K$ users, and $N$ DA ports with independent power supply. For the ease of implementation, both DA ports and users are equipped with single antenna. The TDMA mode is taken into consideration for the downlink multiuser transmission. That is, each frame is divided into $K$ slots, and user $k$ is scheduled in slot $k$ with time duration $\tau_{k}$. We model the channel power gains as $h_{i,k}=cd_{i,k}^{-\phi}\rho^{2}_{i,k}, \forall i,k$, where $c$ is the pathloss at a reference distance of 1m, $d_{i,k}$ denotes the distance between DA port $i$ and user $k$, $\phi$ is the pathloss exponent, and $\rho_{i,k}$ follows independent and identical distribution (i.i.d) with zero mean and unit variance. Note that in DAS, the TDMA transmission for each user $k$ is a multiple-input single-output (MISO) channel and the received signal at user $k$ can be expressed as
\begin{align}\label{eqn:r88}
y_{k}=\bm{g}_{k}^{T}\bm{x}_{k}+z_{k},
\end{align}
where $\bm{g}_{k}=[\sqrt{h_{1,k}}, \ldots, \sqrt{h_{N,k}}]^{T}$ denotes the channel coefficient vector between DA ports and user $k$, $\bm{x}_{k}=[x_{1,k}, \ldots, x_{N,k}]^{T}$ denotes the transmitted signal vector for user $k$, and $z_{k}$ indicates the additive white Gaussian noise (AWGN) with zero mean and variance $\sigma^{2}$. We assume that global channel state information (CSI) is available at the CP. It is also assumed that CSI remains unchanged in each frame but may vary from one frame to another. Denote $\bm{Q}_{k}=E[\bm{x}_{k}\bm{x}_{k}^{\dag}]$ as the covariance of the Gaussian input, where $(\cdot)^{\dag}$ is conjugate transpose of a vector. The achievable rate at user $k$ can be written as
\begin{align}\label{eqn:r89}
R_{k}=\tau_{k}\log\left(1+\frac{1}{\sigma^{2}}\bm{g}_{k}^{T}\bm{Q}_{k}(\bm{g}^{\dag}_{k})^{T}\right).
\end{align}
According to \cite{Vu2011}, in DAS, the DA ports are distributed throughout the area with independent power budget and act independently. Thus the DA ports are without jointly coding and signal processing. Therefore, the transmitted signals at $N$ DA ports are independent and the input covariance is $\bm{Q}_{k}=\text{diag} \{p_{1,k}, \ldots,p_{N, k}\}$, where $p_{i,k}$ is the transmit power between DA port $i$ and user $k$. Then the achievable rate at user $k$ can be further derived as
\begin{eqnarray}
\label{eqn:r02}
R_{k}&=\tau_{k}\log\left(1+\frac{\sum^{N}_{i=1}|g_{i,k}|^{2}p_{i,k}}{\sigma^{2}}\right)\nonumber \\&=\tau_{k}\log\left(1+\frac{\sum^{N}_{i=1}h_{i,k}p_{i,k}}{\sigma^{2}}\right).
\end{eqnarray}
Note that $R_{k}$ in \eqref{eqn:r02} provides a lower bound of the achievable rate of a MISO channel by maximum ratio transmission (MRT). Also it is worth noting that \eqref{eqn:r02} implies DA port selection (or antenna selection/clustering) issue. That is, $p_{i,k}$ is positive if DA port $i$ is selected to transmit information for user $k$, and $p_{i,k}$ should be zero otherwise.

We assume that each user has the energy harvesting function, so that when one particular user $k'$ is scheduled to receive information, the other users $\forall k\neq k'$ can harvest energy from the same RF signal conveying information to user $k'$. Denote $\zeta$ as the energy conversion efficiency, then the harvested energy at user $k$ is given by 
\begin{align}\label{eqn:r01}
E_{k}=\zeta\sum^{N}_{i=1}h_{i,k}\sum^{K}_{k'\neq k}\tau_{k'}p_{i,k'}.
\end{align}
%
%\footnote{Here we only consider the linear EH model with linear energy conversion efficiency $\zeta$. According to \cite{Di2017}, when there is relative low and high power input, the EH receiver works in linear intervals. Moreover, because of signal attenuation, the EH devices are most likely to work under the low input power case, which can be approximately regarded as a linear model. Thus we adopt the linear EH model for the purpose of more tractable analysis in this paper.}%
In \eqref{eqn:r01}, $\sum^{K}_{k'\neq k}\tau_{k'}p_{i,k'}$ is the total energy used for transmitting information from DA port $i$ to all users except user $k$. 
%The noise at each user is modeled as an independent and identically distributed circularly symmetric complex Gaussian noise with variance $\sigma^2$.  Then the achievable rate for user $k$ is 
%
%\begin{align}\label{eqn:r02}
%R_{k}=\tau_{k}\ln\biggl(1+\frac{\sum_{i=1}^N h_{i,k}p_{i,k}}{\sigma^{2}}\biggr).
%\end{align}
%%

Denote $p^{c}_{k}$ as user $k$'s circuit power consumption, like signal processing, mixers, and so on. Then the total energy consumption used for transmitting information to user $k$ can be written as $\tau_{k}(\sum^{N}_{i=1}p_{i,k}+p^{c}_{k}$), which consists of two parts: the energy consumption of transmitting signals via power amplifiers at all DA ports and the circuit energy consumption of user $k$. 
\subsection{UC-EE Maximization Problem Formulation}
From the UC-EE perspective, the EE  of user $k$ is defined as the ratio of its achievable rate and its total energy consumption, which  (in bits/Hz/Joule) can be written as 
\begin{align}\label{eqn:r03}
\eta_{k}=\frac{\tau_{k}\ln\bigl(1+\frac{\sum_{i=1}^N h_{i,k}p_{i,k}}{\sigma^{2}}\bigr)}{\tau_{k}\bigl(\sum^{N}_{i=1}p_{i,k}+p^{c}_{k}\bigr)}=\frac{\ln\bigl(1+\frac{\sum_{i=1}^N h_{i,k}p_{i,k}}{\sigma^{2}}\bigr)}{\sum^{N}_{i=1}p_{i,k}+p^{c}_{k}},
\end{align}
where $\tau_{k}$ is eliminated in the individual EE of user $k$.

The objective of the UC-EE is to balance each user's EE. We adopt the weighted sum EEs of users as the objective function. We maximize the UC-EE of all users by varying the transmit power of DA ports and time duration of each user, subject to the minimum harvested energy requirement $\bar{E}_k$ of each user $k$ and the maximum transmit power constraint $\bar{P_{i}}$ of each DA port $i$. Denote $\bm{p}=\{p_{i,k}\}$ and $\bm{\tau}=\{\tau_{k}\}$, the UC-EE maximization problem can thus be formulated as
\begin{eqnarray}\label{eqn:r04}
 {\rm (P1):}\nonumber
 \max_{\bm{\tau},\bm{p}}&&\sum^{K}_{k=1}w_{k}\eta_{k}  \\
{\rm s.t.}&&E_{k}\ge \bar{E}_k,\forall k,\label{eqn:r05}\\
  && 0\leq \tau_{k}\leq 1,\forall k,\label{eqn:r06}\\
  && 0\leq p_{i,k}\leq \bar{P_{i}},\forall i, k,\label{eqn:r07}\\
  && \sum^{K}_{k=1}\tau_{k}\leq 1,
  \label{eqn:r08}
\end{eqnarray}
where $w_{k}$ is the non-negative constant assigned to user $k$'s EE, denoting user $k$'s EE weight. The weights are parameters decided by the system and reflect the priorities among users.

\subsection{NC-EE Maximization Problem Formulation}
We also consider the EE from the network's perspective. In this case, the network's total energy consumption is 
\begin{align}\label{eqn:r84}
P_{\mathrm{total}} = \sum^{K}_{k=1}\tau_{k}\biggl(\sum^{N}_{i=1}p_{i,k}+p^{c}_{k}\biggr),
\end{align}
which includes the total transmit energy consumption at all DA ports and the total circuit energy consumption at all users. The NC-EE (in bits/Hz/Joule) for this network can be expressed as
\begin{align}\label{eqn:r09}
\eta=\frac{\sum^{K}_{k=1}w_{k}R_{k}}{P_{\mathrm{total}}}.
\end{align}
Similar to the UC-EE problem in (P1),  the NC-EE maximization problem can be formulated as
\begin{eqnarray}\label{eqn:r10}
 {\rm (P2):}\nonumber
 \max_{\bm{\tau},\bm{p}}&&\eta  \\
{\rm s.t.}&&E_{k}\ge \bar{E}_k,\forall k,\label{eqn:r11}\\
  && 0\leq \tau_{k}\leq 1,\forall k,\label{eqn:r12}\\
  && 0\leq p_{i,k}\leq \bar{P_{i}},\forall i,  k,\label{eqn:r13}\\
  && \sum^{K}_{k=1}\tau_{k}\leq 1.
  \label{eqn:r14}
 \end{eqnarray}

\begin{lemma}\label{Lm0}
UC-EE always mathematically outperforms NC-EE.
\end{lemma}

\emph{Proof:}
Please refer to Appendix \ref{AP0}.$\hfill\blacksquare$
\section{Optimal Solution for UC-EE Maximization Problem}\label{se2}
In this section, we address the UC-EE maximization problem (P1). We first transform it into a concave form by some mathematical methods and then develop an iterative algorithm to obtain the globally optimal solution.
\subsection{Problem Transformation}
The constraint \eqref{eqn:r05} is non-convex as two variables are multiplied. To make problem (P1) more tractable, we introduce a set of variables $\bm{s}=\{s_{i,k}\}$ with $s_{i,k}=\tau_{k}p_{i,k}$, which actually denote the energy variables. Then user $k$'s EE $\eta_{k}$ in \eqref{eqn:r03} becomes  
\begin{align}
\eta_{k}=\frac{\ln\biggl(1+\frac{\sum_{i=1}^N h_{i,k}s_{i,k}}{\sigma^{2}\tau_{k}}\biggr)}{\sum^{N}_{i=1}\frac{s_{i,k}}{\tau_{k}}+p^{c}_{k}}=\frac{\tau_{k}\ln\biggl(1+\frac{\sum_{i=1}^N h_{i,k}s_{i,k}}{\sigma^{2}\tau_{k}}\biggr)}{\sum^{N}_{i=1}s_{i,k}+\tau_{k}p^{c}_{k}}.
\end{align}
In addition, substituting $\bm{s}$ into constraints \eqref{eqn:r05} and \eqref{eqn:r07}, problem (P1) can be rewritten as
\begin{eqnarray}
{\rm (P1'):}\nonumber\max_{\bm{\tau},\bm{s}}&&\sum^{K}_{k=1}\frac{w_{k}\tau_{k}\ln\biggl(1+\frac{\sum_{i=1}^N h_{i,k}s_{i,k}}{\sigma^{2}\tau_{k}}\biggr)}{\sum^{N}_{i=1}s_{i,k}+\tau_{k}p^{c}_{k}}\nonumber\\
{\rm s.t.}&&\zeta\sum_{i=1}^{N}h_{i,k}\sum^{K}_{k'\neq k}s_{i,k'} \ge \bar{E}_k,\forall k, \label{C1}\\
  && 0\leq \tau_{k}\leq 1,\forall k,\label{C7} \\
  && 0\leq s_{i,k}\leq \tau_{k}\bar{P_{i}},\forall i, k,\label{C2} \\
  && \sum^{K}_{k=1}\tau_{k}\leq 1.\label{C3}
\end{eqnarray}
Note that once we obtain the optimal energy and time variables $(\bm{s}^*, \bm{\tau}^*)$ by solving problem (P1'), we can recover the optimal power allocation $\bm{p}^*$ by 
\begin{gather}
p_{i,k}^*=\left\{\begin{array}{ll}
s^*_{i,k}/\tau_{k}^* &\text{if } \tau_{k}^*>0,\\
0&
\text{if } \tau^*_{k}=0.\label{eqn:r87}
\end{array}\right.
\end{gather}

The objective function of problem (P1') is with the sum-of-ratios structure and thus non-concave. Based on \cite{Jong2012}, the sum-of-ratios optimization problem can be transformed into a parameterized subtractive-form problem as following.

Denote $\bm{\alpha}=(\alpha_{1},\ldots,\alpha_{K})$ and  $\bm{\beta}=(\beta_{1},\ldots,\beta_{K})$, if $(\bm{\tau}^*,\bm{s}^*)$ is a solution of problem (P1'), there always exists  $\bm{\alpha}^*$ and $\bm{\beta}^*$  such that $(\bm{\tau}^*,\bm{s}^*)$ is also a solution of the following problem with $\bm{\alpha}=\bm{\alpha}^*$ and $\bm{\beta}=\bm{\beta}^*$.
\begin{eqnarray}\label{eqn:r15}
 \max && \sum^{K}_{k=1}\alpha_{k}\biggl(w_{k}R_{k}-\beta_{k}\bigl(\sum^{N}_{i=1}s_{i,k}+\tau_{k}p^{c}_{k}\bigr)\biggr)\nonumber\\
 {\rm s.t.}&& \eqref{C1}, \eqref{C7}, \eqref{C2}, \eqref{C3},
 \end{eqnarray}
where $R_{k}$ can be obtained by applying \eqref{eqn:r87} in \eqref{eqn:r02}. Additionally, $(\bm{\tau}^*,\bm{s}^*)$ also meets the following conditions with  $\bm{\alpha}=\bm{\alpha}^*$ and $\bm{\beta}=\bm{\beta}^*$.
\begin{eqnarray}
 &&1-\alpha_{k}\biggl(\sum^{N}_{i=1}s_{i,k}^*+\tau_{k}^*p^{c}_{k}\biggr)=0, \forall k,\label{eqn:r16}\\
 &&w_{k}R_{k}^*-\beta_{k}\biggl(\sum^{N}_{i=1}s^*_{i,k}+\tau^*_{k}p^{c}_{k}\biggr)=0, \forall k.\label{eqn:r17}
\end{eqnarray}

By the above transformation, now we can solve problem (P1') in an equivalent parameterized form \eqref{eqn:r15} where the objective function is in subtractive form with extra parameters $\bm{\alpha}$ and $\bm{\beta}$. As a result, problem (P1') can be solved in two layers: in the inner-layer, the optimal time and energy variables $(\bm{\tau}^*, \bm{s}^*)$ can be obtained by solving the subtractive formed problem \eqref{eqn:r15} with given $(\bm{\alpha},\bm{\beta})$. At the outer-layer, we find the optimal $(\bm{\alpha}^*,\bm{\beta}^*)$ satisfying \eqref{eqn:r16} and \eqref{eqn:r17}. 

\subsection{Finding Optimal $(\bm{\tau}^*, \bm{s}^*)$ for Given $(\bm{\alpha},\bm{\beta})$}

\begin{lemma}\label{Lm1}
The objective function of problem \eqref{eqn:r15} is jointly concave over $\bm{s}$ and $\bm{\tau}$. 
\end{lemma}

\emph{Proof:}
Please refer to Appendix \ref{AP1}.$\hfill\blacksquare$

Since all constraints in problem \eqref{eqn:r15} are affine, problem \eqref{eqn:r15} is a convex problem and we can use the Lagrangian dual method to solve this substrative formed maximization problem optimally. The Lagrangian function for problem \eqref{eqn:r15} can be written as
\begin{align}\label{eqn:r24}
&L_1(\bm{s},\bm{\tau},\bm{\mu},\bm{\upsilon},\lambda)=\nonumber\\&\sum^{K}_{k=1}\alpha_{k}\biggl(w_{k}\tau_{k}\ln\bigl(1+\frac{\sum_{i=1}^N h_{i,k}s_{i,k}}{\sigma^{2}\tau_{k}}\bigr)-\beta_{k}\bigl(\sum^{N}_{i=1}s_{i,k}+\tau_{k}p^{c}_{k}\bigr)\biggr)\nonumber\\&+\sum^{K}_{k=1}\mu_{k}\biggl(\zeta\sum_{i=1}^{N}h_{i,k}\sum^{K}_{k'\neq k}s_{i,k'}-\bar{E}_k\biggr)+\lambda\biggl(1-\sum^{K}_{k=1}\tau_{k}\biggr)\nonumber\\& +\sum^{K}_{k=1}\sum_{i=1}^{N}\upsilon_{i,k}\biggl(\bar{P_{i}}\tau_{k}-s_{i,k}\biggr),
\end{align}
where $\bm{\mu}=\{\mu_{k}\},$ $\bm{v}=\{v_{i,k}\}$ and $\lambda$ are the Lagrangian multipliers associated with the constraints \eqref{C1},  \eqref{C2} and \eqref{C3} respectively. Then the Lagrangian dual function is given by
\begin{align}\label{eqn:r25}
g_1(\bm{\mu},\bm{\upsilon},\lambda)=\max_{\substack{ \{0\leq \tau_{k}\leq 1\} \\ \{s_{i,k}\geq 0\} }}L_1(\bm{s},\bm{\tau},\bm{\mu},\bm{\upsilon},\lambda).
\end{align}
And the dual problem is given by
\begin{align}\label{eqn:r26}
\min_{\{\bm{\mu}, \bm{\upsilon}, \lambda\}\geq\bm{0} }g_1(\bm{\mu},\bm{\upsilon},\lambda).
\end{align}
%

%\begin{lemma}\label{Lm2}
%With $\sum^{N}_{i=1}s_{i,k}>0$, the optimal time allocation $\tau_{k}^*$ for each user $k$ in \eqref{eqn:r25} should be always positive. 
%\end{lemma}
%
%\emph{Proof:}
%Please refer to Appendix \ref{AP4}.$\hfill\blacksquare$

Now we solve the problem \eqref{eqn:r25} for given  Lagrangian variables $\{\bm{\mu},\bm{\upsilon},\lambda\}$. The BCD method \cite{Richtarik2014} can be adopted to solve the problem, where we alternatively optimize one of $\bm{\tau}$ and $\bm{s}$ with the other fixed. Note that the Lagrangian function $L_{1}$ is jointly concave in $\bm{\tau}$ and $\bm{s}$ as shown before, which ensures the BCD method to converge to the globally optimal solution. 

Given $\bm{s}$, there are two cases of $\tau_{k}^*$. First, if $\sum^{N}_{i=1}s_{i,k}=0$, i.e., $s_{i,k}=0$ for all $i$, which means that there is no energy transmitted for user $k$ and we set $\tau_{k}^{*}=0$ in this case. Otherwise, with $\sum^{N}_{i=1}s_{i,k}>0$, the derivation of the Lagrangian function $L_{1}$ with respect to $\tau_{k}$ is
\begin{align}\label{eqn:r90}
	\frac{\partial L_1}{\partial \tau_{k}}=\alpha_{k}w_{k}\ln\biggl(1+\frac{\sum_{i=1}^N h_{i,k}s_{i,k}}{\sigma^{2}\tau_{k}}\biggr)-\alpha_{k}\beta_{k}p^{c}_{k}-\lambda \nonumber\\-\frac{\alpha_{k}w_{k}\sum^{N}_{i=1}h_{i,k}s_{i,k}}{\sigma^{2}\tau_{k}+\sum^{N}_{i=1}h_{i,k}s_{i,k}}+\sum^{N}_{i=1}\bar{P}_{i}v_{i,k}.
\end{align}
And we can further derive that
\begin{align}\label{eqn:r30}
\frac{\partial^{2} L_1}{\partial \tau^{2}_{k}}=&-\frac{\alpha_{k}w_{k}\bigl(\sum^{N}_{i=1}h_{i,k}s_{i,k}\bigr)^{2}}{\tau_{k}\bigl(\tau_{k}\sigma^{2}+\sum^{N}_{i=1}h_{i,k}s_{i,k}\bigr)^{2}},
\end{align}
%=&-\frac{\alpha_{k}w_{k}\sum^{N}_{i=1}h_{i,k}s_{i,k}}{\tau_{k}^{2}\sigma^{2}+\tau_{k}\sum^{N}_{i=1}h_{i,k}s_{i,k}}+\frac{\sigma^{2}\alpha_{k}w_{k}\sum^{N}_{i=1}h_{i,k}s_{i,k}}{(\tau_{k}\sigma^{2}+\sum^{N}_{i=1}h_{i,k}s_{i,k})^{2}}\nonumber\\
which is negative so that $L_{1}$ is concave over $\tau_{k}$. The solution of $\tau_{k}$ can be solved by setting $\frac{\partial L_1}{\partial \tau_{k}}=0$ as well as considering the constraint $0\leq \tau_{k}\leq 1$. The closed-form solution of $\tau_k^*$ is 
\begin{align}\label{eqn:r91}
	\tau^{*}_{k} =\left[\tilde{\tau}_{k}\right]^{1}_{0},
\end{align}
where $\tilde{\tau}_{k}$ can be obtained by
\begin{gather}
\tilde{\tau}_{k}=\left\{\begin{array}{ll}\label{eqn:r92}
\frac{\sum^{N}_{i=1}h_{i,k}s_{i,k}}{\sigma^{2}(\exp\{\omega(-\exp\{\frac{A_{k}}{\alpha_{k}w_{k}}-1\})+1-\frac{A_{k}}{\alpha_{k}w_{k}}\}-1)}&A_{k}\leq  0,\\
0&
 \text{otherwise,}
\end{array}\right.
\end{gather}
where $A_{k}=\sum^{N}_{i=1}\bar{P}_{i}v_{i,k}-\alpha_{k}\beta_{k}p^{c}_{k}-\lambda$ and $\omega(x)$ is defined as the inverse function of $f(x) = xe^x$ and denotes the principal branch of the Lambert $\omega$ function \cite{Corless1996}. Note that in \eqref{eqn:r92} we have to consider the definition domain of $\omega$ function. The details of obtaining $\tilde{\tau}_{k}$ by solving $\frac{\partial L_1}{\partial \tau_{k}}=0$ can be found at the Appendix B in \cite{Kim2015}.

Next, for given $\bm{\tau}$, by the Karush-Kuhn-Tucker (KKT) conditions \cite{Boyd2004}, we have   
\begin{align}\label{eqn:r27}
\frac{\partial L_1}{\partial s_{i,k}}=&\frac{\alpha_{k}w_{k}\tau_{k}h_{i,k}}{\tau_{k}\sigma^{2}+\sum^{N}_{i=1}s_{i,k}h_{i,k}}-\alpha_{k}\beta_{k}+\zeta\sum^{K}_{k'\neq k}\mu_{k'}h_{i,k'}\nonumber\\&-\upsilon_{i,k}.
\end{align}
By defining $B_{i,k}=\zeta\sum^{K}_{k'\neq k}\mu_{k'}h_{i,k'}-\alpha_{k}\beta_{k}-\upsilon_{i,k}$, the optimal value of $s_{i,k}^*$ can be obtained by setting $\frac{\partial L_1}{\partial s_{i,k}}=0$ since \eqref{eqn:r27} is decreasing with $s_{i,k}$. So $s^*_{i,k}$ can be given by
\begin{align}\label{eqn:r28}
s_{i,k}^*=
\biggl[-\frac{\alpha_{k}w_{k}\tau_{k}}{B_{i,k}}-\frac{\tau_{k}\sigma^{2}+\sum^{N}_{j\neq i}s_{j,k}h_{j,k}}{h_{i,k}}\biggr]^{+},
\end{align}
where $[x]^{+}=\max(x,0)$. Note that we use the BCD method to obtain $s^*_{i,k}$. Thus we update each $s_{i,k}$ as \eqref{eqn:r28} while the values of other $s_{i,k}$'s are given in last iteration.

After solving problem \eqref{eqn:r25} with given $\{\bm{\mu},\bm{\upsilon},\lambda\}$, we now address the minimization problem \eqref{eqn:r26} which is a convex problem. We use the ellipsoid method to simultaneously update $\{\bm{\mu}, \bm{\upsilon}, \lambda\}$ to the optimal ones. The subgradients used for the ellipsoid method are provided as
\begin{align}
&\Delta \mu_k= \zeta\sum^{N}_{i=1}h_{i,k}\sum^{K}_{k'\neq k}s^*_{i,k'}-\bar{E}_{k}, \forall k,\label{eqn:r85}\\
&\Delta \upsilon_{i,k}=\bar{P}_{i}\tau_{k}^* -s^*_{i,k}, \forall i, k,\label{eqn:r82}\\
&\Delta \lambda=1-\sum^{N}_{k=1}\tau^*_{k}.\label{eqn:r81}
\end{align}

\subsection{Finding Optimal $(\bm{\alpha}^*,\bm{\beta}^*)$ for Given $(\bm{\tau}^*, \bm{s}^*)$}
After solving problem \eqref{eqn:r15} with given $\bm{\alpha}$ and $\bm{\beta}$, we now develop an algorithm to update $\bm{\alpha}$ and $\bm{\beta}$ according to \cite{Jong2012}. To begin with, we define $\bm{\psi}(\bm{\alpha, \beta})=(\psi_{1},\ldots,\psi_{2K})$ as
\begin{align}
&\psi_{k}(\alpha_{k})=\alpha_{k}\biggl(\sum^{N}_{i=1}s_{i,k}+\tau_{k}p^{c}_{k}\biggr)-1,\forall k,\label{eqn:r32}\\
&\psi_{k+K}(\beta_{k})=w_{k}R_{k}-\beta_{k}\biggl(\sum^{N}_{i=1}s_{i,k}+\tau_{k}p^{c}_{k}\biggr), \forall k. \label{eqn:r33}
\end{align}
As shown in \cite{Jong2012}, if $\bm{\psi}(\bm{\alpha, \beta})=\bm{0}$, then $(\bm{s}^*, \bm{\tau}^*)$ is the global optimal solution for the problem (P1') and the iteration stops. Otherwise, we need to update $\bm{\alpha}$ and $\bm{\beta}$ as
\begin{align}
&\bm{\alpha}^{n+1}=\bm{\alpha}^{n}+\gamma^{n}\bm{q}^{n},\label{eqn:r34}\\
&\bm{\beta}^{n+1}=\bm{\beta}^{n}+\gamma^{n}\bm{q}^{n},\label{eqn:r35}\\
&\bm{q}^{n}=[\bm{\psi}'(\bm{\alpha, \beta})]^{-1}\bm{\psi}(\bm{\alpha, \beta}),\label{eqn:r36}
\end{align}
where $\bm{\psi}'(\bm{\alpha, \beta})$ is the Jacobian matrix of $\bm{\psi}(\bm{\alpha, \beta})$ and $n$ is the iteration index. Let $m_{k}$ denote the smallest integer among $m\in \{0,1,2,\ldots\}$ which satisfies
\begin{align}\label{eqn:r37}
\lVert \bm{\psi}(\bm{\alpha}^{n}+\xi^{m}\bm{q}^{n}, \bm{\beta}^{n}+\xi^{m}\bm{q}^{n})\rVert \leq (1-\epsilon\xi^{m})\lVert\bm{\psi}(\bm{\alpha, \beta})\rVert,
\end{align}
where $\epsilon\in (0,1)$, $\xi\in (0,1)$, and $\lVert \cdot \rVert$ is the standard Euclidean norm. Then $\gamma^{n}$ can be obtained as $\xi^{m_{k}}$. 

To summarize, the whole algorithm solving problem (P1) optimally is presented in Algorithm \ref{alg:A1}. 

The complexity of Algorithm \ref{alg:A1} is evaluated as follows. The complexity for solving $\bm{s}$ and $\bm{\tau}$ with the BCD method is $\mathcal{O}(K^{2}N)$. The complexity of the ellipsoid method  is $\mathcal{O}((NK+K+1)^{2})$. The complexity for updating $\bm{\alpha}$ and $\bm{\beta}$ is independent of $K$ \cite{Jong2012}.  So the total  complexity of Algorithm \ref{alg:A1} is $\mathcal{O}((NK+K+1)^{2}K^{3}N)$.
\begin{algorithm}[tb]
\caption{Optimal algorithm for problem (P1)  }\label{alg:A1}
\begin{algorithmic}[1]
\STATE Initialize $\bm{\alpha}$ and $\bm{\beta}$.
\REPEAT
\STATE Initialize $\{\bm{\mu}, \bm{\upsilon}, \lambda\}$.
\REPEAT
\STATE Initialize $\bm{s}$ and $\bm{\tau}$.
\REPEAT
\STATE Compute $\tau_{k}$ that maximizes $L_1$ by \eqref{eqn:r91}.
\STATE Compute $\bm{s}$ using \eqref{eqn:r28} with fixed $\bm{\tau}$.
\UNTIL{The improvement of $L_{1}$ stops.}
\STATE Update $\{\bm{\mu}, \bm{\upsilon}, \lambda\}$ by the ellipsoid method using subgradients \eqref{eqn:r85}-\eqref{eqn:r81}.
\UNTIL{$\{\bm{\mu}, \bm{\upsilon}, \lambda\}$ converge to a prescribed accuracy.}
\STATE Denote $m_{k}$ as the smallest $m$ meeting \eqref{eqn:r37}.
\STATE Let $\gamma^{n}=\xi^{m_{k}}$, update $\bm{\alpha}$ and $\bm{\beta}$ by \eqref{eqn:r34} and \eqref{eqn:r35}, respectively.
\UNTIL{$\lVert \bm{\psi}(\bm{\alpha, \beta})\rVert$ is smaller than a prescribed accuracy.} 
\STATE Obtain $\bm{p}^{*}$ by \eqref{eqn:r87}. 
\end{algorithmic}
\end{algorithm}

\section{Optimal Solution for NC-EE Maximization Problem}\label{se4}
In this section, we solve the NC-EE maximization problem (P2). Since problem (P2) is a fractional programming problem, we develop an iterative algorithm to obtain global optimum solution.

To begin with, we define $\mathcal{F}$ as the feasible set of problem (P2) specified by constraints \eqref{eqn:r11}-\eqref{eqn:r14}. Denote $q^{*}$ as the optimal value of problem (P2), we have
\begin{align}\label{eqn:r55}
q^{*}=\max_{(\bm{p},\bm{\tau})\in\mathcal{F}} \frac{\sum^{K}_{k=1}w_{k}R_{k}}{P_{\mathrm{total}}}.
\end{align}
This is a fractional programming problem and we introduce the following theorem, proved in \cite{W.Dinkelbach1967}, to transform this problem into an equivalent linear form. 
\begin{theorem}\label{T1}
 The NC-EE maximization problem (P2) can be solved in the following subtractive form with parameter $q$.
\begin{align}\label{eqn:r56}
\max_{(\bm{p},\bm{\tau})\in\mathcal{F}} \sum^{K}_{k=1}w_{k}R_{k}-qP_{\mathrm{total}}.
\end{align}
Problem \eqref{eqn:r55} and problem \eqref{eqn:r56} are equivalent under the optimal $q^{*}$ if and only if 
\begin{align}\label{eqn:r57}
T(q^{*})=\max_{(\bm{p},\bm{\tau})\in\mathcal{F}} \left\{ \sum^{K}_{k=1}w_{k}R_{k}-q^{*}P_{\mathrm{total}} \right\}=0.
\end{align}
\end{theorem}

From Theorem \ref{T1}, the two problems in \eqref{eqn:r55} and \eqref{eqn:r56} lead to the same optimal solution. Moreover, \eqref{eqn:r57} can be utilized to verify the optimality of the solution in the  subtractive formed problem. 

Based on above, now we can solve problem (P2) optimally with an equivalent form. Here we adopt an iterative algorithm to obtain $q^*$, i.e., the Dinkelbach method  \cite{W.Dinkelbach1967}. In particular, the solution also has a two-layer structure:  for given parameter $q$, we solve the subtractive formed problem in the inner-layer and then we update $q$ by \eqref{eqn:r55} in the outer-layer. This iterative process continues until the optimal solution satisfies condition \eqref{eqn:r57} in Theorem \ref{T1}. The convergence of this algorithm is guaranteed when the subtractive formed problem \eqref{eqn:r56} is solved globally optimally in each iteration.

To make problem \eqref{eqn:r56} more tractable, we set $s_{i,k}=\tau_{k}p_{i,k}$ as in the previous section. As a result, with $\bm{s}=\{s_{i,k}\}$ and given $q$, problem \eqref{eqn:r56} can be reformulated as
\begin{eqnarray}\label{eqn:r58}
 {\rm (P2'):}\max_{\bm{\tau},\bm{s}}&&\sum^{K}_{k=1}w_{k}R_{k}-qP_{\mathrm{total}}  \nonumber\\
{\rm s.t.}&&\zeta\sum^{N}_{i=1}h_{i,k}\sum^{K}_{k'\neq k}s_{i,k'}\ge \bar{E}_k,\forall k,\label{C4}\\
  && 0\leq \tau_{k}\leq 1,\forall k, \\
  && 0\leq s_{i,k}\leq \tau_{k}\bar{P_{i}},\forall i, k, \label{C5}\\
  && \sum^{K}_{k=1}\tau_{k}\leq 1\label{C6}.
 \end{eqnarray}

Similar to the previous section, after solving problem (P2') optimally with $q^{*}$, we can recover the optimal power allocation $\bm{p}^*$ by \eqref{eqn:r87}.

Now problem (P2') is a convex problem and the Lagrangian dual method can be used to solve problem (P2') optimally. The Lagrangian function of problem (P2') for a given $q$ can be written as 
\begin{align}\label{eqn:r59}
&L_2(\bm{s},\bm{\tau},\bm{\mu},\bm{\upsilon},\lambda)=\nonumber\\&\sum^{K}_{k=1}w_{k}\tau_{k}\ln\biggl(1+\frac{\sum_{i=1}^N h_{i,k}s_{i,k}}{\sigma^{2}\tau_{k}}\biggr)-q\sum^{K}_{k=1}\sum^{N}_{i=1}s_{i,k}\nonumber\\&-q\sum^{K}_{k=1}\tau_{k}p^{c}_{k}+\sum^{K}_{k=1}\mu_{k}\biggl(\zeta\sum_{i=1}^{N}h_{i,k}\sum^{K}_{k'\neq k}s_{i,k'}-\bar{E}_k\biggr)\nonumber\\&+\lambda\biggl(1-\sum^{K}_{k=1}\tau_{k}\biggr) +\sum^{K}_{k=1}\sum_{i=1}^{N}\upsilon_{i,k}\biggl(\bar{P_{i}}\tau_{k}-s_{i,k}\biggr),
\end{align}
where $\bm{\mu}=\{\mu_{k}\}$, $\bm{\upsilon}=\{\upsilon_{i,k}\}$ and $\lambda$ are the Lagrangian multipliers with respect to the constraints \eqref{C4}, \eqref{C5} and \eqref{C6}, respectively. Then the corresponding Lagrangian dual function $g_2(\bm{\mu},\bm{\upsilon},\lambda)$ is expressed as
\begin{align}\label{eqn:r60}
g_2(\bm{\mu},\bm{\upsilon},\lambda)=\max_{\substack{ \{0\leq \tau_{k}\leq 1\} \\ \{s_{i,k}\geq 0\} }}L_2(\bm{s},\bm{\tau},\bm{\mu},\bm{\upsilon},\lambda).
\end{align}
The dual problem is written as
\begin{align}\label{eqn:r61}
\min_{\{\bm{\mu}, \bm{\upsilon}, \lambda\}\geq\bm{0} }g_2(\bm{\mu},\bm{\upsilon},\lambda).
\end{align}
Note again that the Lagrangian function $L_{2}$ in \eqref{eqn:r60} is jointly concave in variables $\bm{s}$ and $\bm{\tau}$ as explained in the previous section. Thus we use the BCD method to obtain the optimal solution with the guaranteed convergence. Similar to the previous section, for given $\bm{s}$, we have $\tau^*_{k}=0$ if $s_{i,k}=0$ for all $i$. Otherwise, we have $\frac{\partial^{2} L_2}{\partial \tau^{2}_{k}}<0$ and thus $L_{2}$ is concave over each $\tau_{k}$. So that we solve the zero point of $\frac{\partial L_2}{\partial \tau_{k}}$ within $0\leq \tau_{k}\leq 1$ to obtain the optimal  $\tau_{k}^*$.  We have
\begin{gather}
\tau^*_{k}=\left\{\begin{array}{ll}\label{eqn:r94}
\left[\frac{\sum^{N}_{i=1}h_{i,k}s_{i,k}}{\sigma^{2}(\exp\{\omega(-\exp\{\frac{C_{k}}{w_{k}}-1\})+1-\frac{C_{k}}{w_{k}}\}-1)}\right]^{1}_{0}&C_{k}\leq  0,\\
0&
 \text{otherwise,}
\end{array}\right.
\end{gather}
where $C_{k}=\sum^{N}_{i=1}\bar{P}_{i}v_{i,k}-qp^{c}_{k}-\lambda$. It is worth noting that in \eqref{eqn:r94} the definition domain of $\omega$ function is needed to be taken into consideration as well. And the details of obtaining $\tau^*_{k}$ by solving the zero of $\frac{\partial L_2}{\partial \tau_{k}}$ can be found at the Appendix B in \cite{Kim2015}.

With $\bm{\tau}$ obtained, we can also use the BCD method to optimize $\bm{s}$, i.e., we alternatively optimize each $s_{i, k}$ with the others fixed. The derivation of $L_{2}$ with respect to $s_{i,k}$ can be written as
\begin{align}\label{eqn:r64}
\frac{\partial L_2}{\partial s_{i,k}}=&\frac{w_{k}\tau_{k}h_{i,k}}{\tau_{k}\sigma^{2}+\sum^{N}_{i=1}s_{i,k}h_{i,k}}+D_{i,k},
\end{align}
where $
	D_{i,k}=-q+\zeta\sum^{K}_{k'\neq k}\mu_{k'}h_{i,k'}-\upsilon_{i,k}$.

From \eqref{eqn:r64}, we find that $\frac{\partial L_2}{\partial s_{i,k}}$ is decreasing with $s_{i,k}$. Thus $s_{i,k}^*$ can be uniquely determined through setting $\frac{\partial L_2}{\partial s_{i,k}}=0$ under the non-negative constraint $s_{i,k}\geq 0$.  As a result, to maximize $L_{2}$, we have
\begin{align}\label{eqn:r66}
s_{i,k}^*=
\left[-\frac{w_{k}\tau_{k}}{D_{i,k}}-\frac{\tau_{k}\sigma^2+\sum^{N}_{j\neq i}h_{j,k}s_{j,k}}{h_{i,k}}\right]^{+}.
\end{align}

We also note that for given $\bm{\tau}$, the BCD optimization of $\bm{s}$ by \eqref{eqn:r66} ensures the convergence due to the concavity of $L_2$. 

In summary, problem \eqref{eqn:r60}  can be solved optimally by two BCD loops. In the outer-loop, $\bm{s}$ and $\bm{\tau}$ are alternatively optimized. In the inner-loop, with given $\bm{\tau}$,  each $s_{i,k}$ is also alternatively optimized while fixing other $s_{i,k}$'s. The iterative process stops when the improvement of $L_{2}$ stops. 

Next we turn to obtain the optimal values of Lagrangian multipliers through the ellipsoid method, where the subgradients used to update $\{\bm{\mu}, \bm{\upsilon}, \lambda\}$ are given by
\begin{align}
&\Delta \mu_k=\zeta\sum^{N}_{i=1}h_{i,k}\sum^{K}_{k'\neq k}s^*_{i,k'}-\bar{E}_{k}, \forall k,\label{eqn:r86}\\
&\Delta \upsilon_{i,k}=\bar{P}_{i}\tau_{k}^* -s^*_{i,k}, \forall i, k,\label{eqn:r83}\\
&\Delta \lambda=1-\sum^{N}_{k=1}\tau^*_{k}\label{eqn:r80}.
\end{align}

Finally, after solving the dual function in the pervious steps, we update $q$ as \eqref{eqn:r55}. Then problem (P2') is solved again until $q$ converges to an optimal value $q^*$, which is also the optimal value of NC-EE $\eta^*$. The algorithm for addressing problem (P2) is summarized in Algorithm \ref {alg:A3}.

 The complexity of the BCD method is $\mathcal{O}(K^{2}N)$ and the  complexity of the ellipsoid method is $\mathcal{O}((NK+K+1)^{2})$.  Thus the total  complexity of Algorithm \ref {alg:A3} is $\mathcal{O}(\kappa (NK+K+1)^{2}K^{2}N)$, where $\kappa$ is the number of iterations for updating $q$.

\begin{algorithm}[tb]
\caption{Optimal algorithm for problem (P2)}\label{alg:A3}
\begin{algorithmic}[1]
\STATE Initialize $q$.
\REPEAT
\STATE Initialize $\{\bm{\mu}, \bm{\upsilon}, \lambda\}$.
\REPEAT
\STATE Initialize $\bm{s}$ and $\bm{\tau}$.
\REPEAT
\STATE Compute $\tau_{k}$ that maximizes $L_2$ by \eqref{eqn:r94}.
\STATE Compute $\bm{s}$ using \eqref{eqn:r66} with fixed $\bm{\tau}$.
\UNTIL{The improvement of $L_{2}$ stops.}
\STATE Update $\{\bm{\mu}, \bm{\upsilon}, \lambda\}$ by the ellipsoid method using subgradients \eqref{eqn:r86}-\eqref{eqn:r80}.
\UNTIL{$\{\bm{\mu}, \bm{\upsilon}, \lambda\}$ converge to a prescribed accuracy.}
\STATE Update $q$ as \eqref{eqn:r55}.
\UNTIL{$T(q^{*})$ in \eqref{eqn:r57} is smaller than a prescribed accuracy.}
\STATE Obtain $\bm{p}^{*}$ by \eqref{eqn:r87}.
\end{algorithmic}
\end{algorithm}

\section{Simulation Results}\label{se5}
\begin{table}[t]\label{table:sim_para}
\renewcommand\arraystretch{1.5}
\caption{\\Simulation Parameters} % title of Table
\centering % used for centering table
\begin{tabular}{|c|c|p{'1'}|} % centered columns (2 columns)
\hline %inserts double horizontal lines
Noise power $\sigma^{2}$ & $-$104 dBm \\  % inserts table heading
\hline % inserts single horizontal line
Pathloss at a reference distance of 1m & $10^{-3}$ \\
\hline % inserts single horizontal line
Pathloss exponent & 2 \\
\hline 
Length of the square & 10 m \\ %
\hline  % inserting body of the table
Power constraint for the $i$-th DA port & $\bar{P_{i}}=\bar{P}$\\ %
\hline  % inserting body of the table
Harvested energy constraint for the $k$-th user & $\bar{E}_k=\bar{E}$\\ %
\hline  % inserting body of the table
Circuit power consumption & $p^{c}_{k}=0.5 \text{W}, \forall k$ \\
\hline 
Weight of users & $w_{k}=1,\forall k$ \\ %
\hline 
DA port deployment & Square layout \\ %
\hline 
Energy conversion efficiency $\zeta$ & 0.6 \\ %
%\hline 
%Number of channel realizations & 1000 \\ %
%\hline 
%Number of device generations & 1000 \\ %
\hline %inserts single line
\end{tabular}
\label{table:sim_para} % is used to refer this table in the text
\end{table}

In this section, we present simulation results to verify the effectiveness of the proposed optimal algorithms for UC-EE and NC-EE maximization problems called UC-OPT and NC-OPT, respectively. The main system parameters are listed in Table \ref{table:sim_para}. In the proposed DAS, we assume that $N$ DA ports are distributed uniformly in a square area of  $100$ square meters and the users are  randomly distributed throughout the area.  For comparison, we also evaluate the performance of the following benchmark schemes: 

\begin{enumerate}
	\item \textbf{UC-EE maximization problem with fixed time allocation (UC-FT)}. In this scheme, the information transmission time for each user is fixed as $\tau_{k}=1/K, \forall k$.  Note that this is a special case of problem (P1) and the proposed  Algorithm \ref{alg:A1} is also applicable for this case. The overall complexity for this benchmark scheme is $\mathcal{O}(K^{4}N)$.
	\item \textbf{UC-EE maximization problem with fixed power allocation (UC-FP)}. The transmit power in each DA port is fixed as $p_{i,k}=\bar{P}_{i}, \forall i, k$ in this case. As a result, the user's EE becomes $\eta_{k}=\frac{\ln\bigl(1+\frac{\sum_{i=1}^N h_{i,k}\bar{P}_{i}}{\sigma^{2}}\bigr)}{\sum^{N}_{i=1}\bar{P}_{i}+p^{c}_{k}}$ which is a constant. Thus in this case we need to optimize the transmit time $\bm{\tau}$ to meet the minimum harvested energy constraints. 
	\item \textbf{NC-EE maximization problem with fixed time allocation (NC-FT)}. With $\tau_{k}=1/K, \forall k$,  Algorithm \ref{alg:A3} is also applicable for solving this simplified NC-EE maximization problem. The total complexity is $\mathcal{O}(\kappa K^{3}N)$.
	\item \textbf{NC-EE maximization problem with fixed power allocation (NC-FP)}. Given $p_{i,k}=\bar{P}_{i}, \forall i, k$, we adopt the Dinkelbach method to transform  this time allocation problem into a linear programming problem. Therefore, we can apply some standard linear optimization methods, such as the simplex method \cite{Boyd2004}, to obtain the solution efficiently. The total complexity is $\mathcal{O}(\kappa K)$.
\end{enumerate}
\begin{figure}[t]
\begin{centering}
\includegraphics[scale=0.4]{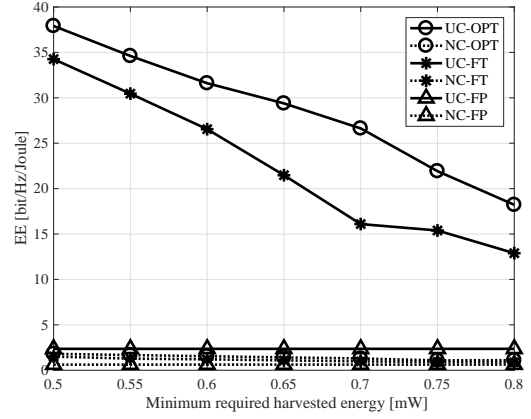}
\vspace{-0.1cm}
 \caption{  UC-EE and NC-EE versus the minimum harvested energy constraint $\bar{E}$. }\label{fig:E_min}
\end{centering}
\vspace{-0.1cm}
\end{figure}

Fig. \ref{fig:E_min} illustrates the impact of the minimum harvested energy requirement on the EE of the considered schemes with 4 users, 7 DA ports, and $\bar{P}=6 \text{W}$. From the figure, the optimality of the proposed schemes (UC-OPT and NC-OPT) is confirmed. Also when the minimum harvested energy constraint $\bar{E}$ increases, the EE performance for all considered schemes declines for two reasons. First, each user needs longer time for energy harvesting to meet the growing minimum harvested energy demand. This results in shorter time for information decoding, which finally results in a lower throughput. Secondly, with increasing $\bar{E}$, the DA ports are likely to transmit higher power so as to enable the users to harvest more energy, which leads to higher energy consumption. Moreover, we observe that the benchmark scheme with fixed time allocation outperforms the benchmark scheme with fixed power allocation, no matter in the UC case or NC case. It demonstrates that power allocation plays a more important role in the optimization process, compared with time allocation. Furthermore, we note that the UC-OPT scheme gains much more EE than the NC-OPT scheme. The reason for this performance gap is that the UC-OPT scheme adopts the weighted sum EEs of individual users as its performance metric, which is in sum-of-ratios structure, while the NC-OPT scheme chooses the ratio of system's throughput to total energy consumption as its performance metric, which is in fractional structure.   
\begin{figure}[t]
\begin{centering}
\includegraphics[scale=0.4]{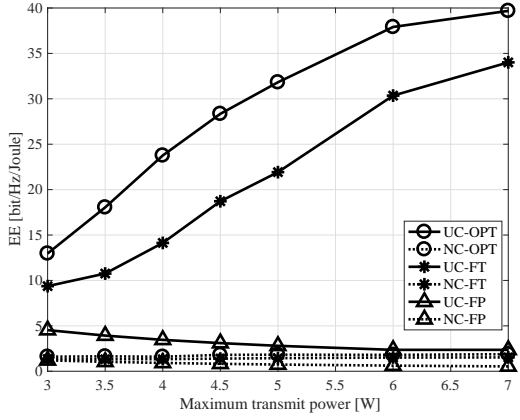}
\vspace{-0.1cm}
 \caption{  UC-EE and NC-EE versus the maximum transmit power constraint $\bar{P}$. }\label{fig:P_max}
\end{centering}
\vspace{-0.1cm}
\end{figure}

In Fig. \ref{fig:P_max}, we compare the EE performance of the above mentioned schemes with respect to the maximum transmit power constraint $\bar{P}$. The numbers of DA ports and users are 7 and 4 respectively, and $\bar{E}$ is fixed as $0.1 \text{mW}$ in this case. First, we confirm the effectiveness of our proposed optimal schemes (UC-OPT and NC-OPT). We observe that as the maximum transmit power constraint grows, the UC-OPT, NC-OPT, UC-FT and NC-FT schemes' EE increase. In particular, the UC-OPT and UC-FT schemes experience a sharp increase in their EE first, but gradually saturate in the high transmit power region. This is because the minimum harvested energy requirement $\bar{E}$ is relatively high when $\bar{P}$ is small, which means that there are strict requirements and few network resources. Thus a small increase in $\bar{P}$ can have a significant improvement of the EE performance. For $\bar{P}$ in high region, we have rich network resources to meet the minimum harvested energy demand. Therefore, there is a high flexibility in resource allocation and $\bar{P}$ no longer has a large impact on EE performance. However, with the augment of $\bar{P}$, the UC-FP and NC-FP schemes' EE decline.  Note that the transmit power is fixed as $\bar{P}$ in these two benchmark schemes so that the numerators of the individual EE $\eta_{k}$ in \eqref{eqn:r03} and NC-EE $\eta$ in \eqref{eqn:r09} have a logarithmic growth with increasing $\bar{P}$ while the denominators increase linearly. Hence the gap between the optimal schemes (UC-OPT and NC-OPT) and the benchmark schemes with fixed transmit power (UC-FP and NC-FP) widens in the high $\bar{P}$ region.

\begin{figure}[t]
\begin{centering}
\includegraphics[scale=0.4]{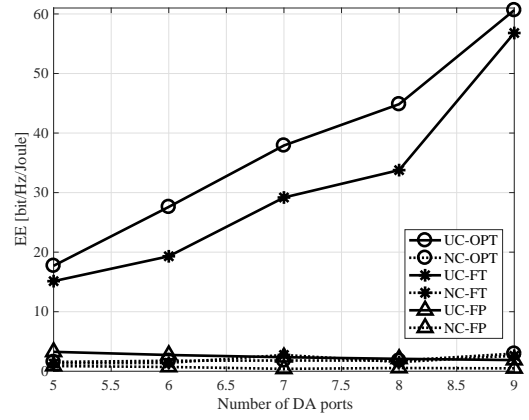}
\vspace{-0.1cm}
 \caption{  UC-EE and NC-EE versus the number of DA ports $N$. }\label{fig:Number_DA}
\end{centering}
\vspace{-0.1cm}
\end{figure}

 In Fig. \ref{fig:Number_DA}, we show the relationship between the number of DA ports and the EE of the considered schemes with 4 users and $\bar{E}=0.2 \text{mW}$. The UC-OPT gains most UC-EE, compared with other UC-EE benchmark schemes (UC-FT, UC-FP) and so does the case of the NC-EE maximization schemes. Note that with the growing number of DA ports, the EE performance of all schemes improves as expected. With more DA ports, i.e., more network resources, we have high flexibility in time allocation and power allocation. Moreover, since the DA ports are uniformly distributed in a fixed area, more DA ports mean shorter average access distances between the users and DA ports. Both reasons lead to better EE performance. However, as for the benchmark schemes with fixed power allocation, we observe a modest decrease in their EE performance with the growth of the number of DA ports. This is because in these two benchmark schemes, all DA ports are turned on and transmit with maximum power $\bar{P}$. As a result, the individual EE $\eta_{k}$ in \eqref{eqn:r03} and the NC-EE $\eta$ in \eqref{eqn:r09} experience a linear increase in the denominator while the nominator has a logarithmic increase. 

 \begin{figure}[t]
\begin{centering}
\includegraphics[scale=0.4]{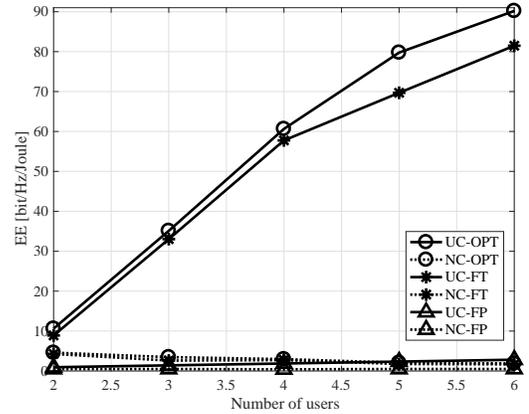}
\vspace{-0.1cm}
 \caption{ UC-EE and NC-EE versus the number of users. }\label{fig:user}
\end{centering}
\vspace{-0.1cm}
\end{figure}

The impact of the number of users on the considered schemes is demonstrated in Fig. \ref{fig:user}. As can be seen,  the EE performance of the UC-EE maximization schemes (UC-OPT, UC-FT and UC-FP) improve with more users. This is because that the UC-EE maximization schemes all choose the weighted sum EEs of users as their objective functions which increases with the growing number of users. In particular, we observe that the UC-EE maximization schemes' EE tend to be saturated as the number of users increases. The reason accounting for this trend is that each user has minimum harvested energy requirement in this system. When there are more users, i.e., there are limited network resources and growing demand of overall minimum harvested energy requirements, the improvement of UC-EE finally saturates. This reason is also applicable for the decrease in the NC-EE maximization schemes (NC-OPT, NC-FT and NC-FP).

 \begin{figure}[t]
\begin{centering}
\includegraphics[scale=0.4]{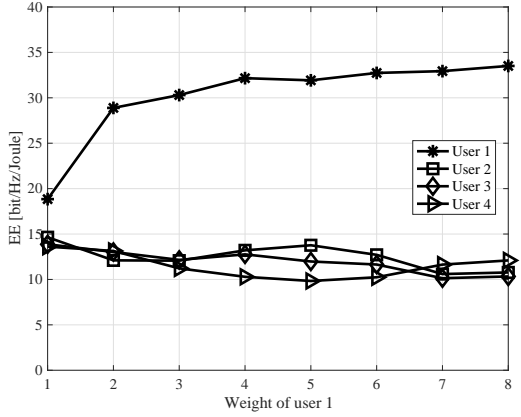}
\vspace{-0.1cm}
 \caption{  UC-EE versus the weight of user 1. }\label{fig:weight}
\end{centering}
\vspace{-0.1cm}
\end{figure}

Fig. \ref{fig:weight} illustrates the EE tradeoff between four users where users 2, 3 and 4, are assigned the same weights, i.e., $w_{2}=w_{3}=w_{4}=1$ while the weight of user 1, $w_{1}$ is varied between 1 and 8. There are 7 DA ports and $\bar{E}=0.2 \text{mW}$. From Fig. \ref{fig:weight}, we can see that with growing $w_{1}$, the EE of user 1 shows an upward trend while the EEs of other users fall. Based on this trend, we can obtain that the improvement of user's EE performance can be achieved by assigning higher weight, which also means a flexibility of customizing the EE performance of different users. Specially, with increasing $w_{1}$, the EE of user 1  experiences a considerable increase first but finally approaches a maximum value, indicating that assigning higher weight to user has a limit effect on improving its EE performance. 

\section{Conclusions}\label{se6}
In this paper, we have investigated energy-efficient resource allocation in WIPT based DAS. Two kinds of EE metrics have been studied, namely, NC-EE and UC-EE.  We have formulated the UC-EE and NC-EE maximization problems where the transmit power and time are jointly optimized.  As both problems are nonlinear programming problems and thus non-convex, we have proposed iterative algorithms to find the optimal solutions by using some mathematical transformations.  

Some valuable insights have been provided through extensive simulations: First, UC-EE always outperforms NC-EE. Second, power allocation is more important than time allocation for improving EE. Third, more users benefit UC-EE but harm NC-EE.

\begin{appendices}
\section{Proof of Lemma \ref{Lm0}} \label{AP0}
 Firstly we define a set of new variables $T_{k}, \forall k$ as $T_{k}=\tau_{k}(\sum^{N}_{i=1}p_{i,k}+p^{c}_{k}), \forall k$ which indicates the total energy consumption of transmitting information to user $k$. To avoid confusion, for the NC-EE maximization problem, we denote the optimal values of $T_{k}$ and $R_{k}$ as $T^{NC}_{k}$ and $R^{NC}_{k}$ respectively, $\forall k$ . Similarly, in terms of the UC-EE maximization problem, we denote the optimal values of $T_{k}$ and $R_{k}$ as $T^{UC}_{k}$ and $R^{UC}_{k}$ respectively, $\forall k$. Then we have
\begin{align}\label{eqn:r22}	
\eta=\frac{\sum_{k=1}^{K}{w_{k}R^{NC}_{k}}}{\sum_{k=1}^{K}{T^{NC}_{k}}}\leq \max_{k}\frac{w_{k}R^{NC}_{k}}{T^{NC}_{k}}\leq\sum_{k=1}^{K} \frac{w_{k}R^{NC}_{k}}{T^{NC}_{k}}.
\end{align}
Because $T^{NC}_{k}$ and $R^{NC}_{k}$, $\forall k$ are the optimal values of the NC-EE maximization problem, for the UC-EE maximization problem with the same constraints, we can further derive that    
\begin{align}\label{eqn:r29}	
\sum_{k=1}^{K} \frac{w_{k}R^{NC}_{k}}{T^{NC}_{k}}\leq \sum_{k=1}^{K} \frac{w_{k}R^{UC}_{k}}{T^{UC}_{k}}=\sum_{k=1}^{K}w_{k}\eta_{k}.
\end{align}
From \eqref{eqn:r29} and \eqref{eqn:r22} we can conclude that UC-EE always outperforms NC-EE. 
\section{Proof of Lemma \ref{Lm1}} \label{AP1}
With $\bm{s}$, user $k$'s achievable rate $R_{k}$ becomes
\begin{align}
	R_{k}=\tau_{k}\ln\biggl(1+\frac{\sum^{N}_{i=1}h_{i,k}s_{i,k}}{\sigma^{2}\tau_{k}}\biggr).
\end{align}
To keep $R_{k}$ continuity over $0\leq\tau_{k}\leq1, \forall k$, here we define $R_{k}=0$ when $\tau_{k}=0$ for all $k$. 

The objective function of problem \eqref{eqn:r15} can be written as $\sum^{K}_{k=1}\alpha_{k}f_{1}(\bm{s},\tau_{k})$, where 
\begin{align}
f_1(\bm{s},\tau_{k})=\left\{\begin{array}{ll}\label{eqn:r23}
w_{k}R_{k} -\beta_{k}\biggl(\sum^{N}_{i=1}s_{i,k}+\tau_{k}p^{c}_{k}\biggr)&\tau_{k}>0,\\
-\beta_{k}\sum^{N}_{i=1}s_{i,k}&
\tau_{k}=0.
\end{array}\right.
\end{align}
According to \cite{Boyd2004}, to prove the concavity of $\sum^{K}_{k=1}\alpha_{k}f_{1}(\bm{s},\tau_{k})$, we need to prove that for $(\hat{\bm{s}},\hat{\tau}_{k})=\theta(\dot{\bm{s}},\dot{\tau}_{k})+(1-\theta)(\ddot{\bm{s}},\ddot{\tau}_{k}), 0<\theta< 1$, $f_1(\hat{\bm{s}},\hat{\tau}_{k})\geq\theta f_1(\dot{\bm{s}},\dot{\tau}_{k})+(1-\theta)f_1(\ddot{\bm{s}},\ddot{\tau}_{k})$ is always satisfied. Here four mutually complementary cases for $\dot{\tau}_{k}$ and $\ddot{\tau}_{k}$ are considered by us. 
\begin{enumerate}
  \item $\dot{\tau}_{k}>0$ and $\ddot{\tau}_{k}>0$:  In this case, $\hat{\tau}$ is also positive. According to \cite{Boyd2004},  $w_{k}\tau_{k}\ln\biggl(1+\frac{\sum_{i=1}^N h_{i,k}s_{i,k}}{\sigma^{2}\tau_{k}}\biggr)$ is jointly concave over $\bm{s}$ and $\tau_{k}$. Then we can further derive that $f_1(\bm{s},\tau_{k})=w_{k}\tau_{k}\ln\biggl(1+\frac{\sum_{i=1}^N h_{i,k}s_{i,k}}{\sigma^{2}\tau_{k}}\biggr) -\beta_{k}\biggl(\sum^{N}_{i=1}s_{i,k}+\tau_{k}p^{c}_{k}\biggr)$ is also jointly concave over $\bm{s}$ and $\tau_{k}$. As a result, we have $f_1(\hat{\bm{s}},\hat{\tau}_{k})\geq\theta f_1(\dot{\bm{s}},\dot{\tau}_{k})+(1-\theta)f_1(\ddot{\bm{s}},\ddot{\tau}_{k})$ in this case.
  \item $\dot{\tau}_{k}>0$ and $\ddot{\tau}_{k}=0$: In this case,  $f_1(\ddot{\bm{s}},\ddot{\tau}_{k})=-\beta_{k}\sum^{N}_{i=1}\ddot{s}_{i,k}$ and $f_1(\hat{\bm{s}},\hat{\tau}_{k})$ can be expressed as 
  \begin{align}\label{eqn:r24}
&f_{1}(\hat{\bm{s}},\hat{\tau}_{k})=-\beta_{k}\theta\bigl(\sum^{N}_{i=1}\dot{s}_{i,k}+\dot{\tau}_{k}p^{c}_{k}\bigr)-\beta_{k}(1-\theta)\sum^{N}_{i=1}\ddot{s}_{i,k}\nonumber\\&+\!\theta \dot{\tau}_{k}w_{k}\ln\biggl(\!1\!+\!\frac{\sum_{i=1}^N h_{i,k}\dot{s}_{i,k}}{\sigma^{2}\dot{\tau}_{k}}\!+\!\frac{(1\!-\!\theta)\sum_{i=1}^N h_{i,k}\ddot{s}_{i,k}}{\sigma^{2}\theta\dot{\tau}_{k}}\!\biggr).
\end{align}
	So $f_1(\hat{\bm{s}},\hat{\tau}_{k})\geq\theta f_1(\dot{\bm{s}},\dot{\tau}_{k})+(1-\theta)f_1(\ddot{\bm{s}},\ddot{\tau}_{k})$ is proved in this case.
  \item $\dot{\tau}_{k}=0$ and $\ddot{\tau}_{k}>0$: Since this case is similar to the second case, we can draw the same conclusion based on the previous analysis.
  \item $\dot{\tau}_{k}=0$ and $\ddot{\tau}_{k}=0$: In this case,  $f_1(\dot{\bm{s}},\dot{\tau}_{k})$ and $f_1(\ddot{\bm{s}},\ddot{\tau}_{k})$ equal to $-\beta_{k}\sum^{N}_{i=1}\dot{s}_{i,k}$ and $-\beta_{k}\sum^{N}_{i=1}\ddot{s}_{i,k}$, respectively. Noting that they are both linear functions. Therefore, $f_1(\hat{\bm{s}},\hat{\tau}_{k})\geq\theta f_1(\dot{\bm{s}},\dot{\tau}_{k})+(1-\theta)f_1(\ddot{\bm{s}},\ddot{\tau}_{k})$ is satisfied.
\end{enumerate}
\end{appendices}

 \begin{footnotesize}
\bibliographystyle{IEEEtran}
 \bibliography{JSAC-1570440702}

\begin{thebibliography}{10}
\providecommand{\url}[1]{#1}
\csname url@samestyle\endcsname
\providecommand{\newblock}{\relax}
\providecommand{\bibinfo}[2]{#2}
\providecommand{\BIBentrySTDinterwordspacing}{\spaceskip=0pt\relax}
\providecommand{\BIBentryALTinterwordstretchfactor}{4}
\providecommand{\BIBentryALTinterwordspacing}{\spaceskip=\fontdimen2\font plus
\BIBentryALTinterwordstretchfactor\fontdimen3\font minus
  \fontdimen4\font\relax}
\providecommand{\BIBforeignlanguage}[2]{{%
\expandafter\ifx\csname l@#1\endcsname\relax
\typeout{** WARNING: IEEEtran.bst: No hyphenation pattern has been}%
\typeout{** loaded for the language `#1'. Using the pattern for}%
\typeout{** the default language instead.}%
\else
\language=\csname l@#1\endcsname
\fi
#2}}
\providecommand{\BIBdecl}{\relax}
\BIBdecl

\bibitem{Lu2015}
X.~Lu, P.~Wang, D.~Niyato, D.~I. Kim, and Z.~Han, ``Wireless networks with {RF}
  energy harvesting: A contemporary survey,'' \emph{IEEE Commun. Surveys
  Tuts.}, vol.~17, no.~2, pp. 757--789, 2015.

\bibitem{Ju2014c}
H.~Ju and R.~Zhang, ``Throughput maximization in wireless powered communication
  networks,'' \emph{IEEE Trans. Wireless Commun.}, vol.~13, no.~1, pp.
  418--428, Jan. 2014.

\bibitem{Clerckx2017}
B.~Clerckx, Z.~B. Zawawi, and K.~Huang, ``Wirelessly powered backscatter
  communications: Waveform design and {SNR}-energy tradeoff,'' \emph{IEEE
  Commun. Lett.}, vol.~21, no.~10, pp. 2234--2237, Oct. 2017.

\bibitem{Ju2014a}
H.~Ju and R.~Zhang, ``Optimal resource allocation in full-duplex
  wireless-powered communication network,'' \emph{IEEE Trans. Commun.},
  vol.~62, no.~10, pp. 3528--3540, Oct. 2014.

\bibitem{Liu2016}
Y.~Liu and X.~Wang, ``Information and energy cooperation in {OFDM} relaying:
  Protocols and optimization,'' \emph{IEEE Trans. Veh. Technol.}, vol.~65,
  no.~7, pp. 5088--5098, Jul. 2016.

\bibitem{Zhang2013}
R.~Zhang and C.~K. Ho, ``{MIMO} broadcasting for simultaneous wireless
  information and power transfer,'' \emph{IEEE Trans. Wireless Commun.},
  vol.~12, no.~5, pp. 1989--2001, May 2013.

\bibitem{Liu2016a}
Y.~Liu, ``Wireless information and power transfer for multirelay-assisted
  cooperative communication,'' \emph{IEEE Commun. Lett.}, vol.~20, no.~4, pp.
  784--787, Apr. 2016.

\bibitem{Liu2013}
L.~Liu, R.~Zhang, and K.~C. Chua, ``Wireless information and power transfer: A
  dynamic power splitting approach,'' \emph{IEEE Trans. Commun.}, vol.~61,
  no.~9, pp. 3990--4001, Sep. 2013.

\bibitem{Zhang2016c}
M.~Zhang, Y.~Liu, and R.~Zhang, ``Artificial noise aided secrecy information
  and power transfer in {OFDMA} systems,'' \emph{IEEE Trans. Wireless Commun.},
  vol.~15, no.~4, pp. 3085--3096, Apr. 2016.

\bibitem{Liu2017a}
Y.~Liu, ``Joint resource allocation in {SWIPT}-based multiantenna
  decode-and-forward relay networks,'' \emph{IEEE Trans. Veh. Technol.},
  vol.~66, no.~10, pp. 9192--9200, Oct. 2017.

\bibitem{Clerckx2018}
B.~Clerckx, ``Wireless information and power transfer: Nonlinearity, waveform
  design, and rate-energy tradeoff,'' \emph{IEEE Trans. Signal Process.},
  vol.~66, no.~4, pp. 847--862, Feb. 2018.

\bibitem{Liu2017}
M.~Liu and Y.~Liu, ``Power allocation for secure {SWIPT} systems with
  wireless-powered cooperative jamming,'' \emph{IEEE Commun. Lett.}, vol.~21,
  no.~6, pp. 1353--1356, Jun. 2017.

\bibitem{Zhou2003}
S.~Zhou, M.~Zhao, X.~Xu, J.~Wang, and Y.~Yao, ``Distributed wireless
  communication system: a new architecture for future public wireless access,''
  \emph{IEEE Commun. Mag.}, vol.~41, no.~3, pp. 108--113, Mar. 2003.

\bibitem{hu2007}
H.~Hu, Y.~Zhang, and J.~Luo, \emph{Distributed Antenna Systems: Open
  Architecture for Future Wireless Communications}.\hskip 1em plus 0.5em minus
  0.4em\relax CRC Press, 2007.

\bibitem{He2013}
C.~He, B.~Sheng, P.~Zhu, X.~You, and G.~Y. Li, ``Energy- and
  spectral-efficiency tradeoff for distributed antenna systems with
  proportional fairness,'' \emph{IEEE J. Sel. Areas Commun.}, vol.~31, no.~5,
  pp. 894--902, May 2013.

\bibitem{Li2016}
X.~Li, X.~Ge, X.~Wang, J.~Cheng, and V.~C.~M. Leung, ``Energy efficiency
  optimization: Joint antenna-subcarrier-power allocation in {OFDM}-{DAS}s,''
  \emph{IEEE Trans. Wireless Commun.}, vol.~15, no.~11, pp. 7470--7483, Nov.
  2016.

\bibitem{Ng2015}
D.~W.~K. Ng and R.~Schober, ``Secure and green {SWIPT} in distributed antenna
  networks with limited backhaul capacity,'' \emph{IEEE Trans. Wireless
  Commun.}, vol.~14, no.~9, pp. 5082--5097, Sep. 2015.

\bibitem{Yuan2015}
F.~Yuan, S.~Jin, Y.~Huang, K.~k.~Wong, Q.~T. Zhang, and H.~Zhu, ``Joint
  wireless information and energy transfer in massive distributed antenna
  systems,'' \emph{IEEE Commun. Mag.}, vol.~53, no.~6, pp. 109--116, Jun. 2015.

\bibitem{Yuan2017}
F.~Yuan, S.~Jin, K.~K. Wong, J.~Zhao, and H.~Zhu, ``Wireless information and
  power transfer design for energy cooperation distributed antenna systems,''
  \emph{IEEE Access}, vol.~5, pp. 8094--8105, 2017.

\bibitem{Zhang2017}
S.~Zhang, Q.~Wu, S.~Xu, and G.~Y. Li, ``Fundamental green tradeoffs:
  Progresses, challenges, and impacts on {5G} networks,'' \emph{IEEE Commun.
  Surveys Tuts.}, vol.~19, no.~1, pp. 33--56, 2017.

\bibitem{Miao2013}
G.~Miao, ``Energy-efficient uplink multi-user {MIMO},'' \emph{IEEE Trans.
  Wireless Commun.}, vol.~12, no.~5, pp. 2302--2313, May 2013.

\bibitem{Ng2013a}
D.~W.~K. Ng, E.~S. Lo, and R.~Schober, ``Energy-efficient resource allocation
  in {OFDMA} systems with hybrid energy harvesting base station,'' \emph{IEEE
  Trans. Wireless Commun.}, vol.~12, no.~7, pp. 3412--3427, Jul. 2013.

\bibitem{Xiong2014}
C.~Xiong, L.~Lu, and G.~Y. Li, ``Energy efficiency tradeoff in downlink and
  uplink {TDD} {OFDMA} with simultaneous wireless information and power
  transfer,'' in \emph{Proc. IEEE Int. Conf. Commun. (ICC)}, Jun. 2014, pp.
  5383--5388.

\bibitem{Ng2013}
D.~W.~K. Ng, E.~S. Lo, and R.~Schober, ``Wireless information and power
  transfer: Energy efficiency optimization in {OFDMA} systems,'' \emph{IEEE
  Trans. Wireless Commun.}, vol.~12, no.~12, pp. 6352--6370, Dec. 2013.

\bibitem{Huang2018a}
Y.~Huang, M.~Liu, and Y.~Liu, ``Energy-efficient {SWIPT} in {IoT} distributed
  antenna systems,'' \emph{IEEE Internet Things J.}, vol.~PP, no.~99, p.~1,
  2018.

\bibitem{Yu2015}
G.~Yu, Q.~Chen, R.~Yin, H.~Zhang, and G.~Y. Li, ``Joint downlink and uplink
  resource allocation for energy-efficient carrier aggregation,'' \emph{IEEE
  Trans. Wireless Commun.}, vol.~14, no.~6, pp. 3207--3218, Jun. 2015.

\bibitem{Yu2015a}
G.~Yu, Y.~Jiang, L.~Xu, and G.~Y. Li, ``Multi-objective energy-efficient
  resource allocation for multi-{RAT} heterogeneous networks,'' \emph{IEEE J.
  Sel. Areas Commun.}, vol.~33, no.~10, pp. 2118--2127, Oct. 2015.

\bibitem{Wu2016b}
Q.~Wu, W.~Chen, D.~W.~K. Ng, J.~Li, and R.~Schober, ``User-centric energy
  efficiency maximization for wireless powered communications,'' \emph{IEEE
  Trans. Wireless Commun.}, vol.~15, no.~10, pp. 6898--6912, Oct. 2016.

\bibitem{Ding2018}
J.~Ding, L.~Jiang, and C.~He, ``User-centric energy-efficient resource
  management for time switching wireless powered communications,'' \emph{IEEE
  Commun. Lett.}, vol.~22, no.~1, pp. 165--168, Jan. 2018.

\bibitem{Vu2011}
M.~Vu, ``{MISO} capacity with per-antenna power constraint,'' \emph{IEEE Trans.
  Commun.}, vol.~59, no.~5, pp. 1268--1274, May 2011.

\bibitem{Jong2012}
Y.-C. Jong, ``An efficient global optimization algorithm for nonlinear
  sum-of-ratios problem,'' [Online]. Available:
  \url{http://www.optimization-online.org/DB_FILE/2012/08/3586.pdf}.

\bibitem{Richtarik2014}
P.~Richtárik and M.~Takáč, ``Iteration complexity of randomized
  block-coordinate descent methods for minimizing a composite function,''
  \emph{Math. Program.}, vol. 144, pp. 1--38, Dec. 2011.

\bibitem{Corless1996}
R.~M. Corless, G.~H. Gonnet, D.~E.~G. Hare, D.~J. Jeffrey, and D.~E. Knuth,
  ``On the {LambertW} function,'' \emph{Adv. Comput. Math.}, vol.~5, pp.
  329--359, 1996.

\bibitem{Kim2015}
H.~Kim, S.~R. Lee, C.~Song, K.~J. Lee, and I.~Lee, ``Optimal power allocation
  scheme for energy efficiency maximization in distributed antenna systems,''
  \emph{IEEE Trans. Commun.}, vol.~63, no.~2, pp. 431--440, Feb. 2015.

\bibitem{Boyd2004}
S.~Boyd and L.~Vandenberghe, \emph{Convex optimization II}.\hskip 1em plus
  0.5em minus 0.4em\relax Cambridge University Press, 2004.

\bibitem{W.Dinkelbach1967}
W.~Dinkelbach, ``On nonlinear fractional programming,'' \emph{Manage. Sci.},
  vol.~13, no.~9, pp. 492--498, Mar. 1967.

\end{thebibliography}
\end{footnotesize}
\begin{IEEEbiography}
[{\includegraphics[width=1in,height=1.25in,clip,keepaspectratio]{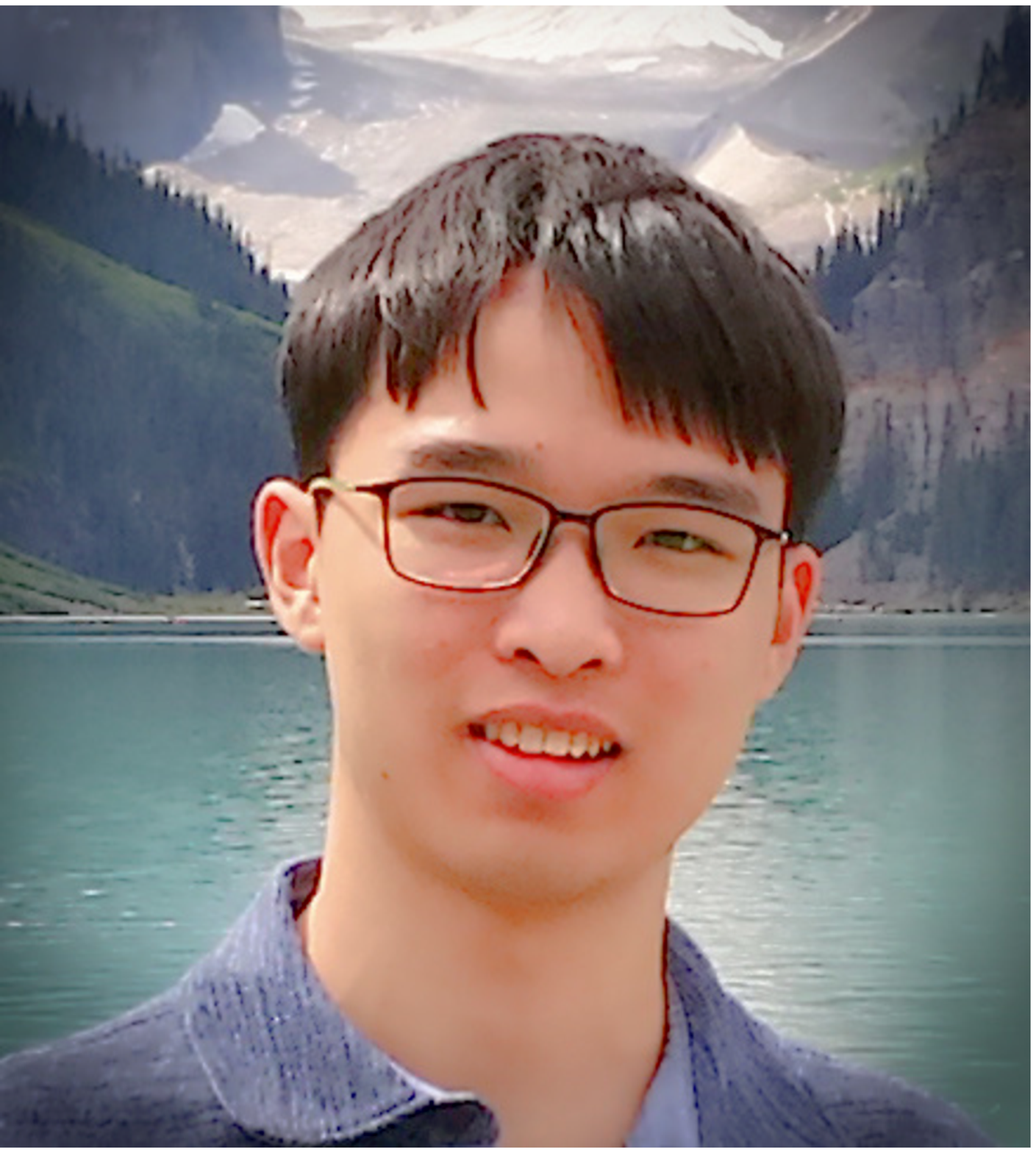}}]
{Yuwen Huang} received the B.S. degree from South China University of Technology, Guangzhou, China, in 2018. He is currently working towards his Ph.D degree in the Department of Information Engineering at The Chinese University of Hong Kong (CUHK). His research interests lie in wireless communications and information theory.
\end{IEEEbiography} 
\begin{IEEEbiography}
[{\includegraphics[width=1in,height=1.25in,clip,keepaspectratio]{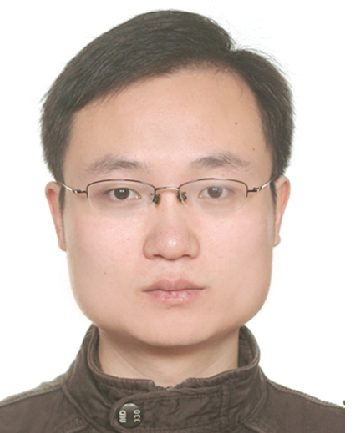}}]
{Yuan Liu}(S'11-M'13'-SM'18)
received the B.S. degree from Hunan University of Science and Technology, Xiangtan, China, in 2006; the M.S. degree from Guangdong University of Technology, Guangzhou,
China, in 2009; and the Ph.D. degree from Shanghai Jiao Tong University, China, in 2013, all in electronic engineering.
Since Fall 2013, he has been with the School of Electronic and Information Engineering, South China University of Technology, Guangzhou, where he is currently an associate professor.

Dr. Liu serves as an editor for the \textsc{IEEE Communications Letters} and the \textsc{IEEE Access}. His research interests include 5G communications and beyond, mobile edge computation offloading, and machine learning in wireless networks.\end{IEEEbiography} 

\begin{IEEEbiography}
[{\includegraphics[width=1.5in,height=1.25in,clip,keepaspectratio]{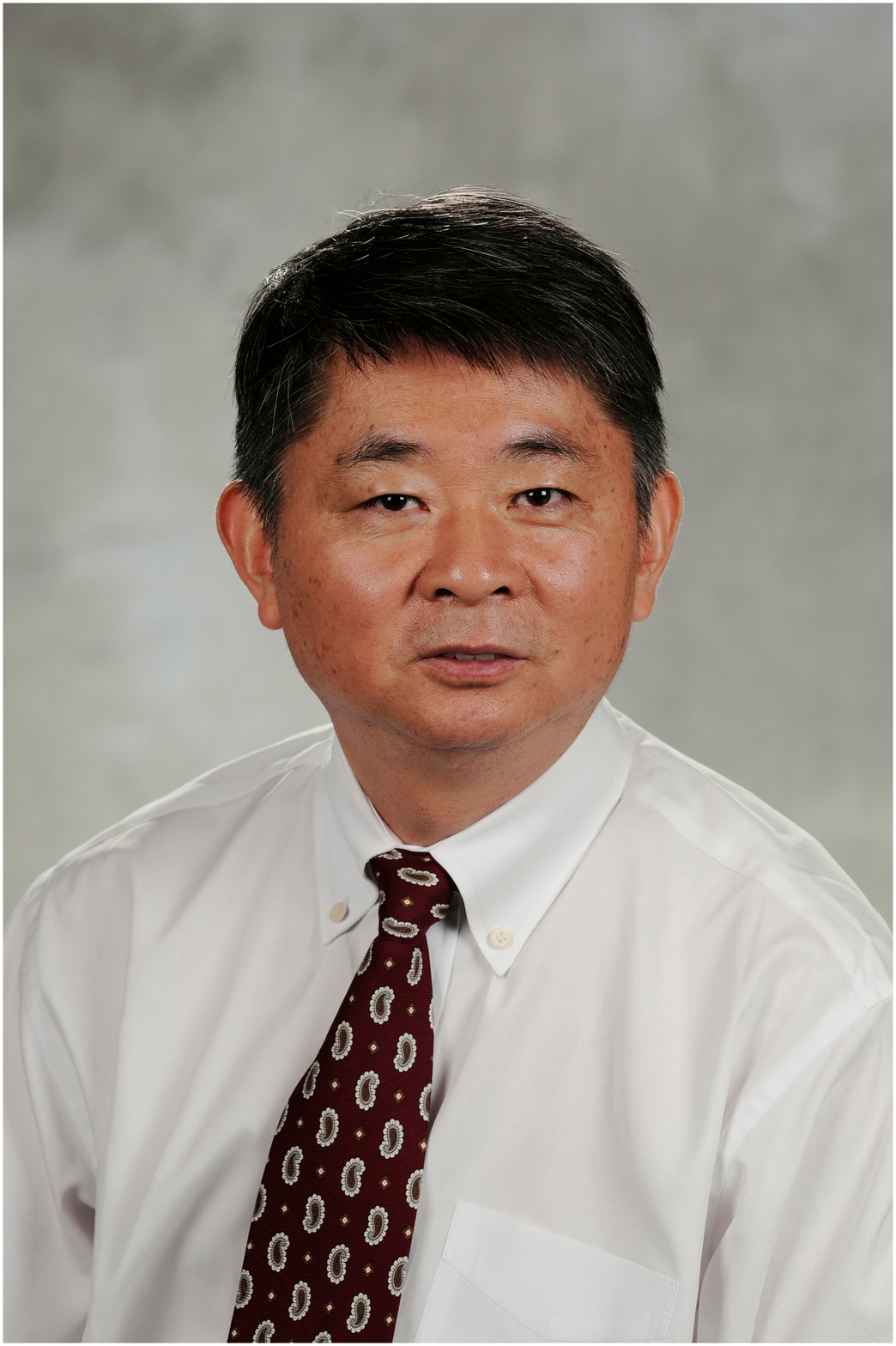}}]
{Geoffrey Ye Li} (S'93-M'95-SM'97-F'06)  received his B.S.E. and M.S.E. degrees in 1983 and 1986, respectively, from the Department of Wireless Engineering, Nanjing Institute of Technology, Nanjing, China, and his Ph.D. degree in 1994 from the Department of Electrical Engineering, Auburn University, Alabama. 

He was a Teaching Assistant and then a Lecturer with Southeast University, Nanjing, China, from 1986 to 1991, a Research and Teaching Assistant with Auburn University, Alabama, from 1991 to 1994, and a Post-Doctoral Research Associate with the University of Maryland at College Park, Maryland, from 1994 to 1996. He was with AT\&T Labs - Research at Red Bank, New Jersey, as a Senior and then a Principal Technical Staff Member from 1996 to 2000. Since 2000, he has been with the School of Electrical and Computer Engineering at the Georgia Institute of Technology as an Associate Professor and then a Full Professor. He is also holding a Cheung Kong Scholar title at the University of Electronic Science and Technology of China since 2006.  

His general research interests include statistical signal processing and machine learning for wireless communications. In these areas, he has published over 200 journal papers in addition to over 40 granted patents and many conference papers. His publications have been cited over 32,000 times and he has been recognized as the \textit{World's Most Influential Scientific Mind}, also known as a \textit{Highly-Cited Researcher}, by Thomson Reuters almost every year. He was awarded \textit{IEEE Fellow} for his contributions to \textit{signal processing for wireless communications} in 2005. He won 2010 \textit{IEEE ComSoc Stephen O. Rice Prize Paper Award}, 2013 \textit{IEEE VTS James Evans Avant Garde Award}, 2014 \textit{IEEE VTS Jack Neubauer Memorial Award}, 2017 \textit{IEEE ComSoc Award for Advances in Communication}, and 2017 \textit{IEEE SPS Donald G. Fink Overview Paper Award}. He also received \textit{2015 Distinguished Faculty Achievement Award} from the \textit{School of Electrical and Computer Engineering, Georgia Tech}. He has been involved in editorial activities for over 20 technical journals for the IEEE, including founding Editor-in-Chief of \textit{IEEE 5G Tech Focus}. He has organized and chaired many international conferences, including technical program vice-chair of \textit{IEEE ICC'03}, technical program co-chair of \textit{IEEE SPAWC'11}, general chair of \textit{IEEE GlobalSIP'14}, and technical program co-chair of \textit{IEEE VTC'16 (Spring)}. 
\end{IEEEbiography}

%\begin{IEEEbiography}
%[{\includegraphics[width=1in,height=1.25in,clip,keepaspectratio]{Yuan_Liu_photo.eps}}]
%{Yuan Liu}(S'11-M'13)
%received the B.S. degree from Hunan University of Science and Technology, Xiangtan, China, in 2006; the M.S. degree from Guangdong University of Technology, Guangzhou,
%China, in 2009; and the Ph.D. degree from Shanghai Jiao Tong University, China, in 2013, all in electronic engineering.
%
%Since Fall 2013, he has been an Assistant Professor with South China University of Technology, Guangzhou. His current research interests include heterogeneous networks, cooperative relay communication, and physical-layer security.
%
%Dr. Liu is the recipient of the Guangdong Province Excellent Master Theses Award in 2010. He has been honored as an Exemplary Reviewer of the \textsc{IEEE Communications Letters}. He is also awarded the IEEE Student Travel Grant for IEEE ICC 2012.
%\end{IEEEbiography}

\end{document}